\documentclass[conference]{IEEEtran}
\usepackage{ifpdf}
\usepackage{cite}
\usepackage{algorithmic}
\usepackage{graphicx}
\usepackage{amsmath}
\usepackage{url}
\graphicspath{{fig}{images}}
\DeclareGraphicsExtensions{.ps,.eps,.pdf,.jpeg,.png}
\begin{document}
%
% paper title
% can use line breaks \\ within to get better formatting as desired
\title{Programming Massively Parallel Architectures using MARTE: a Case Study}

% author names and affiliations
% use a multiple column layout for up to three different
% affiliations
\author{\IEEEauthorblockN{A. Wendell Rodrigues}
\IEEEauthorblockA{}
\and
\IEEEauthorblockN{Fr\'{e}d\'{e}ric Guyomarc'h}
\IEEEauthorblockA{LIFL - USTL\\\small{INRIA Lille Nord Europe} - 59650\\Villeneuve d'Ascq - France\\\{wendell.rodrigues, frederic.guyomarch, jean-luc.dekeyser\}@inria.fr}
\and
\IEEEauthorblockN{Jean-Luc Dekeyser}
\IEEEauthorblockA{}
}

% make the title area
\maketitle

\begin{abstract}
%\boldmath

Nowadays, several industrial applications are being ported to parallel architectures. These applications take advantage of the potential parallelism provided by multiple core processors. Many-core processors, especially the GPUs(Graphics Processing Unit), have led the race of floating-point performance since 2003. While the performance improvement of general-purpose microprocessors has slowed significantly, the GPUs have continued to improve relentlessly. As of 2009, the ratio between many-core GPUs and multicore CPUs for peak floating-point calculation throughput is about 10 times. However, as parallel programming requires a non-trivial distribution of tasks and data, developers find it hard to implement their applications effectively. Aiming to improve the use of many-core processors, this work presents an case-study using UML and MARTE profile to specify and generate OpenCL code for intensive signal processing applications. Benchmark results show us the viability of the use of MDE approaches to generate GPU applications.

\end{abstract}
\IEEEpeerreviewmaketitle

\section{Introduction}
Advanced engineering and scientific communities have used parallel programming to solve their large scale complex problems for a long time. Despite the high level knowledge of the developers belonging to these communities, they find hard to implement their parallel applications effectively. Over recent years, using Graphics Processing Units
(GPUs) has become increasingly popular and in fact important for seeking performance benefits in computationally intensive parallel applications.
However, even for GPUs, there are not performance gains without challenges: first, the identification and exploitation of any parallelism in the application is responsibility of programmers. Often, this requires intimate understanding of the hardware and extensive re-factoring work rather than simple program transformations. Second, the high-level abstractions of a problem can hardly be expressed in the CUDA\cite{cuda} or OpenCL\cite{opencl10} programming model. Furthermore, subsequent manual optimisations distort any remaining abstractions in the application.

A way of addressing this abstraction is to provide a \emph{model-to-source} transformation mechanism where the model is captured through a Model-Driven Environment (MDE)\cite{rodrigues11} and then the code generation is handled by templates. Gaspard2 \cite{gaspard2} is a framework that uses UML and the MARTE profile in order to implement its MDE approach. This paper shows a case-study using Gaspard2 to generate an application code for massively parallel architectures based on GPUs. Moreover, performance issues are applied to the model transformation and code generation steps in order to achieve optimization levels accomplished in manually written codes.

\section{Background}
\subsection{Massively Parallel Architectures}
GPU is a manycore processor attached to a graphics card dedicated to calculating floating point operations. The GPU devotes more transistors to data processing rather data caching and flow control. This is the reason why the GPU is specialized for compute intensive. Nevertheless, even if GPUs are a manycore processors, their parallelism continues to scale with Moore's law. It is necessary to develop application software that transparently scales its parallelism. GPUs such as NVIDIA GeForce GTX 480 contain 15 Streaming Multiprocessors, each of which supports up to 1024 co-resident threads, so 30K threads can be created for a certain task. In addition, each multiprocessor executes groups, called warps, of 32 threads simultaneously.
NVIDIA's actual CUDA architecture, code-named Fermi, has features for general-purpose computing. Fundamentally, Fermi processors are still graphics processors, not general-purpose processors. The system still needs a host CPU to run the operating system, supervise the GPU, provide access to main memory, present a user interface, and perform everyday tasks that have little or no data-level parallelism.

\subsection{OpenCL}
OpenCL(Computing Language) is a standard for parallel computing consisting of a language, API, libraries and a runtime system. Originally, was proposed by Apple, and then turned over to the Khronos Group. 

OpenCL also defines a programming language for writing
kernels, which is an extension of C. Kernels are executed
within their own memory domain and may not directly
access host main memory. OpenCL usually defines a host where main programs are placed and one or more devices that executes kernels. Furthermore, device memory is divided into
four distinct regions:
\begin{itemize}
\item \textit{Global memory}, a kind of "device main memory",
can be accessed by all work-items and the host in
reads/writes.
\item \textit{Constant memory} is similar to global memory, except
that work-items may only read from this memory.
\item \textit{Local memory} is read/write memory local to a work-
group, and is shared by all work-items of this group.
\item \textit{Private memory} is local to each work-item.
\end{itemize}
The OpenCL programming language defines type qualifiers to specify in which memory region a
variable is stored or a pointer points to. As a kernel can neither access host main memory nor dynamically allocate global and constant memory, all memory management must be done by the host. The OpenCL API provides functions to allocate linear memory blocks in global or constant memory, as well as to copy data to or from these blocks.

\subsection{MARTE in Gaspard2 Context}
MARTE (Modeling and Analysis of Real-Time and Embedded systems) \cite{marte10} is a standard proposal of the Object Management Group (OMG).
The primary aim of MARTE is to add capabilities to UML for model-driven engineering of real-time and embedded systems. 
UML provides the framework into which needed concepts are plugged. The MARTE profile enhances possibility to model software, hardware and relations between
them. It also provides extensions to make performance and scheduling analysis and to take into account platform services. Gaspard2 \cite{gaspard2} is a framework based on MDE and MARTE profile. From a high-level abstraction model of application, architecture and allocation, Gaspard2 provides transformation chains and templates to code generation for several target platforms. One of these platforms is the hybrid (CPU+GPU) platform beneath the OpenCL API.

\section{Case Study: H.263 Video Downscaling}
\label{sec:caseproblem}

The case study concerned in this paper deals with specific aspect of
H.263-based video compression standard, scaling.  The scaling during
video-compression is considerably important for previews or for
streaming for small form factor devices, such as mobile phones.  The
application consists of a classical downscaler, which transforms a
video signal, which, for instance, is expressed in Common Intermediate
Format (CIF), into a smaller size video.  In this situation, the
downscaler can be composed of two components: a horizontal filter that
reduces the number of pixels from 352-lines to 132-lines and a
vertical filter that reduces the number of pixels from 288-lines to
128-lines by interpolating packets of 8 pixels both row- and
column-wise.

\begin{figure}[!h]
\begin{center}
\includegraphics[width=.47\textwidth]{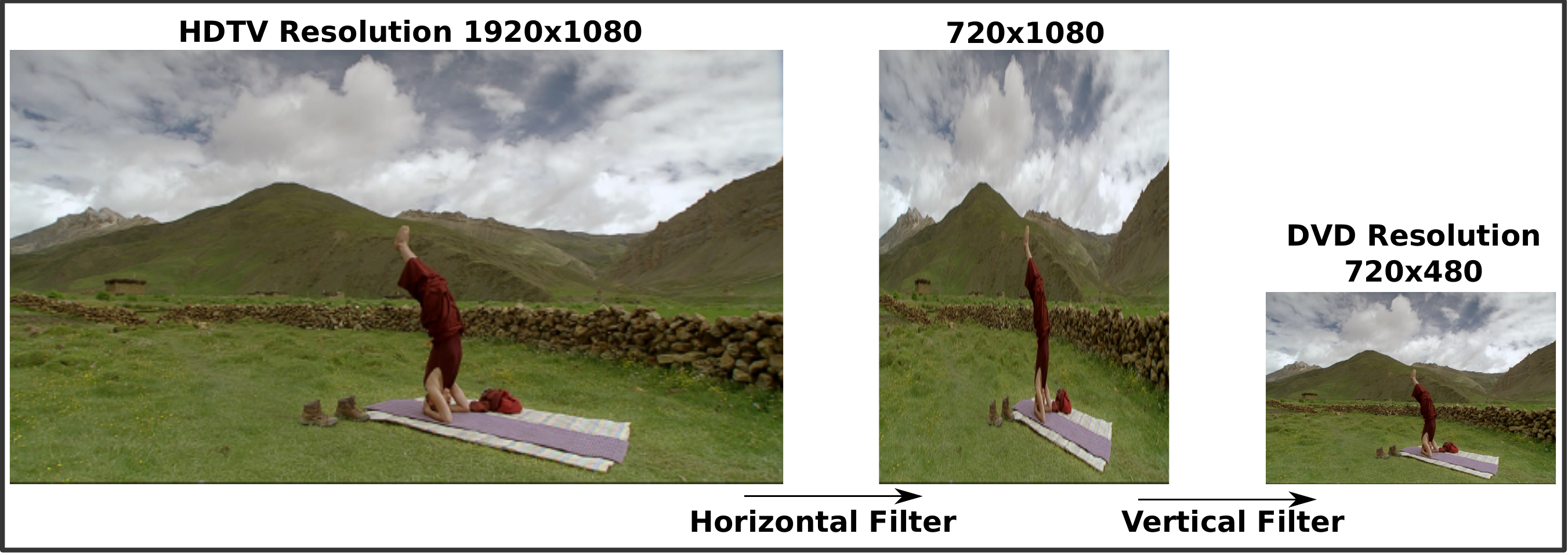}
\end{center}
\vspace*{-.4cm}
\caption{Horizontal and vertical filter processes}
\label{fig:video-compression}
\end{figure}

In a typical case of handling a 25-frames-per-second video signal
lasting for 80-seconds, the downscaler may process up to 2000 frames in
CIF format, with each input frame being represented by a
two-dimensional array of size $352\times 288$ and should emit $2000$
output frames of size $132\times 128$.  Since each video pixels is
encoded in 24-bit RGB colour model, the frame generation process is
repeated for each frame and for each pixel of different colour space
along two different directions. The final frame is produced by using
these outputs from different colour space. Depending on the composing
function, a broad-range of output colours are possible for each pixel
and thus for each frame. The figure~\ref{fig:video-compression}
illustrates this basic operation for a given frame in high-definition
format.

As can be observed, the operations concerned with the scaling is
highly parallel and repetitive. The interpolation is repeated for
each frame, each pixel and for each colour channel.

\subsection{Downscaler Model}
\begin{figure}[!h]
\begin{center}
\includegraphics[width=.45\textwidth]{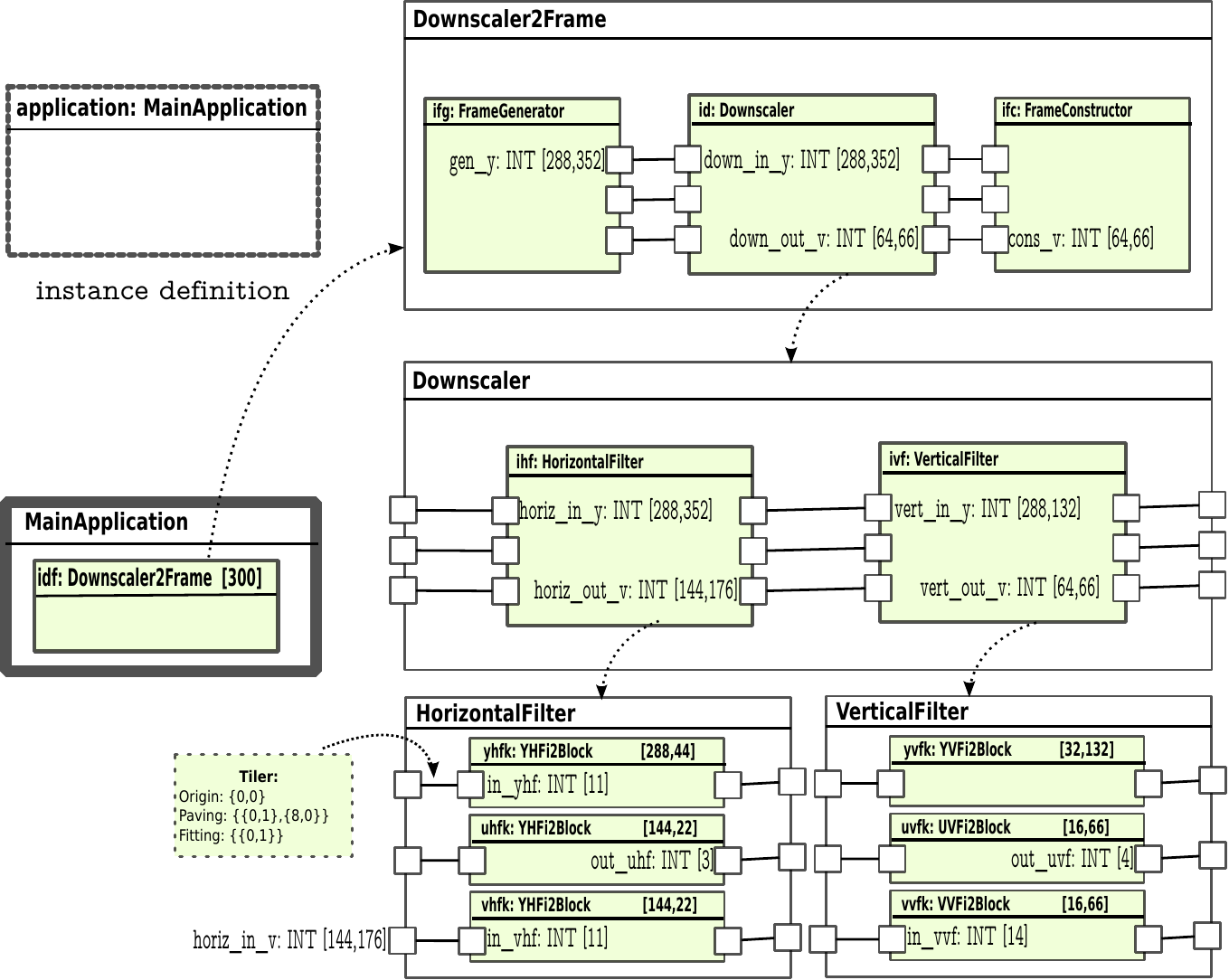}
\end{center}
\caption{Downscaler Application Model}
\label{fig-dscalerapp}
\end{figure}

The figure \ref{fig-dscalerapp} gives us an overview of the downscaler application. The figure illustrates a model for a 300-frames video, even if we have 2000 frames in our testbed. For this example, we are going to analyse only the first repetitive task from the Horizontal Filter component. The other tasks have equivalent behaviour. The \emph{yhfk} task has a multiplicity equals to [288,44]. It means this task is composed of 288x44 independent tasks (so-called Elementary Task), and thus, parallelizable. Each elementary task takes a pattern from the input array. A \emph{tiler} stereotype\footnote{defined in ArrayOL\cite{bouletaol} language and part of MARTE.} do the tiling operation. It allows to split input data in patterns in accordance with \emph{tiler}'s array definitions of \emph{origin, paving and fitting}. Besides features such as tiler specifications and task repetitions, MARTE profile is applied to OpenCL architecture definition (host and device) in order to grant task allocations. For this illustrated model, we add a specification to one host (CPU+Memory) and one device (GPU+Global Memory). Data and tasks are placed into memories and processors according to project interests. For instance, the six repetitive tasks in \textit{horizontal} and \textit{vertical} filters are allocated onto the GPU in order to generate kernels in the execution environment. Allocate stereotypes are used to map ports and tasks into \textit{HwRAMs} and \textit{HwProcessors}. These stereotypes will allow for creating all variables and relations between them. Additionally, in order to distinguish host from device, we modify the description attribute from \textit{HwResource} stereotype.

\subsection{MARTE to OpenCL Transformation Chain}
Gaspard2 supplies a transformation engine that allows us chaining a set of model-to-model or model-to-text transformations. These transformations take into account model elements and properties and gather information to create cleaner models towards a target platform. A recent chain was added to Gaspard2 framework and it allows for automatic code generation from MARTE to OpenCL (see \cite{rodrigues11}). Subsequent paragraphs highlight some details of the designed model taken into account by these transformations.
\subsubsection{Launch Topology}
\begin{figure}[!h]
\begin{center}
\includegraphics[width=.40\textwidth]{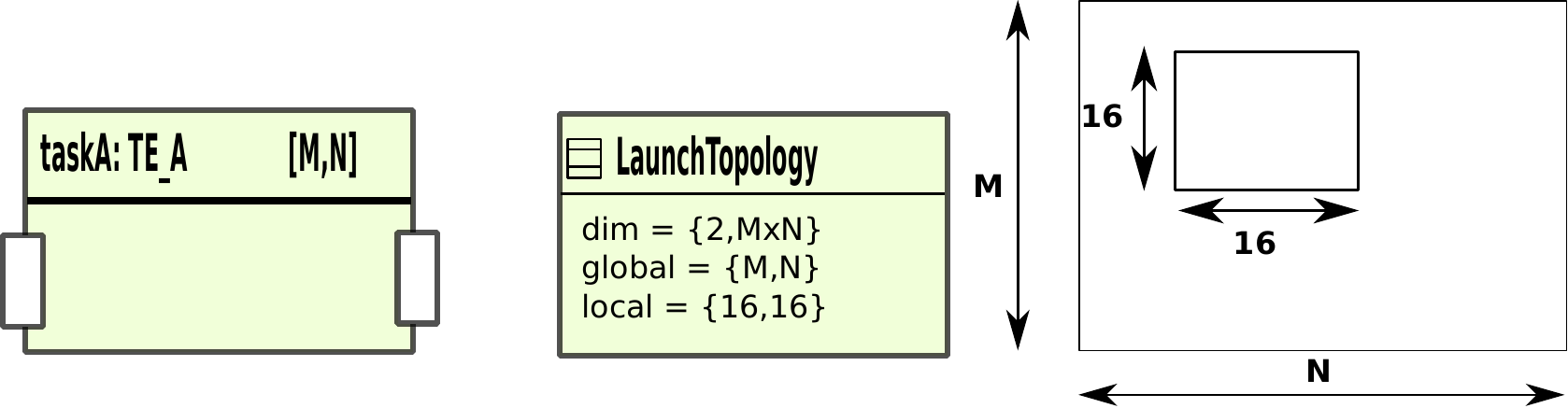}
\end{center}
\caption{Downscaler Application Model}
\label{fig-ltopology}
\end{figure}

An allocated repetitive task should be properly executed. In OpenCL programming model, elementary tasks are work-items in a work-group context. The work-group and work-item topology (grid of threads) are computed from multiplicity of the elementary task. For instance, a MxN multiplicity is transformed in the work-item topology as defined in the figure \ref{fig-ltopology}. Threshold levels help to avoid mistaken topology definitions for smaller or bigger multiplicities.

\subsubsection{Data Allocation}
A critical problem in application modeling based on MDE is to manage the data allocation in the target platform. MARTE profile adds the \textit{flowPort} stereotype to UML \textit{port} element. The main attribute aggregated to a port element is the \textit{direction}, which allows to define whether the port is input, output, or bidirectional. This information contributes to decide which elements are read-only variables.

\begin{figure}[!h]
\begin{center}
\includegraphics[width=.45\textwidth]{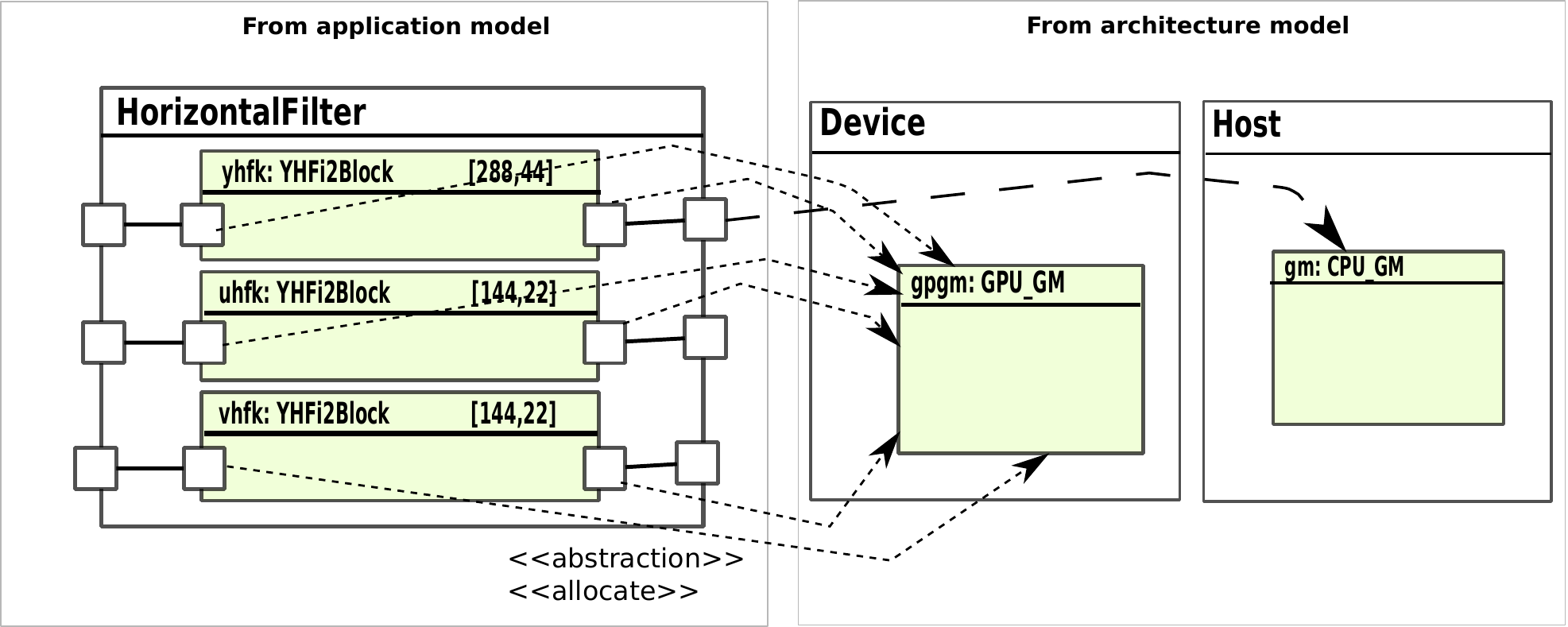}
\end{center}
\caption{Downscaler Application Model}
\label{fig-memoryalloc}
\end{figure}

By using UML links we can associate \textit{flowPorts} to memories in architecture models. Each port has attributes and associations that allow us defining size and data type for example. Thus, developers can specify in their application models where the data will be stored and how much space will occupy the data. In the figure \ref{fig-memoryalloc} we can see a simple example of allocation. Ports from different tasks are allocated into memory elements of their respective processors.

\subsubsection{Performance Tuning}
Usually connected ports allocated to different memory boxes, as seen in the data allocations illustration (figure \ref{fig-memoryalloc}), cause a data transfer between CPU and GPU. At a first sight, one can say that is a critical point to decrease the performance because subsequent kernels reuse these data. Since unnecessary data transfer times are expressive in running time, it would be interesting to take out execessive transfers in the model design. Nevertheless, applying some inteligency levels to the transformations we can detect these critical points in the original model and avoid extra data transfers. Therefore, performance gains, as observed in result charts in next section, can be achieved automatically.
\section{Results}
\begin{figure}[!h]
\begin{center}
\includegraphics[width=.50\textwidth]{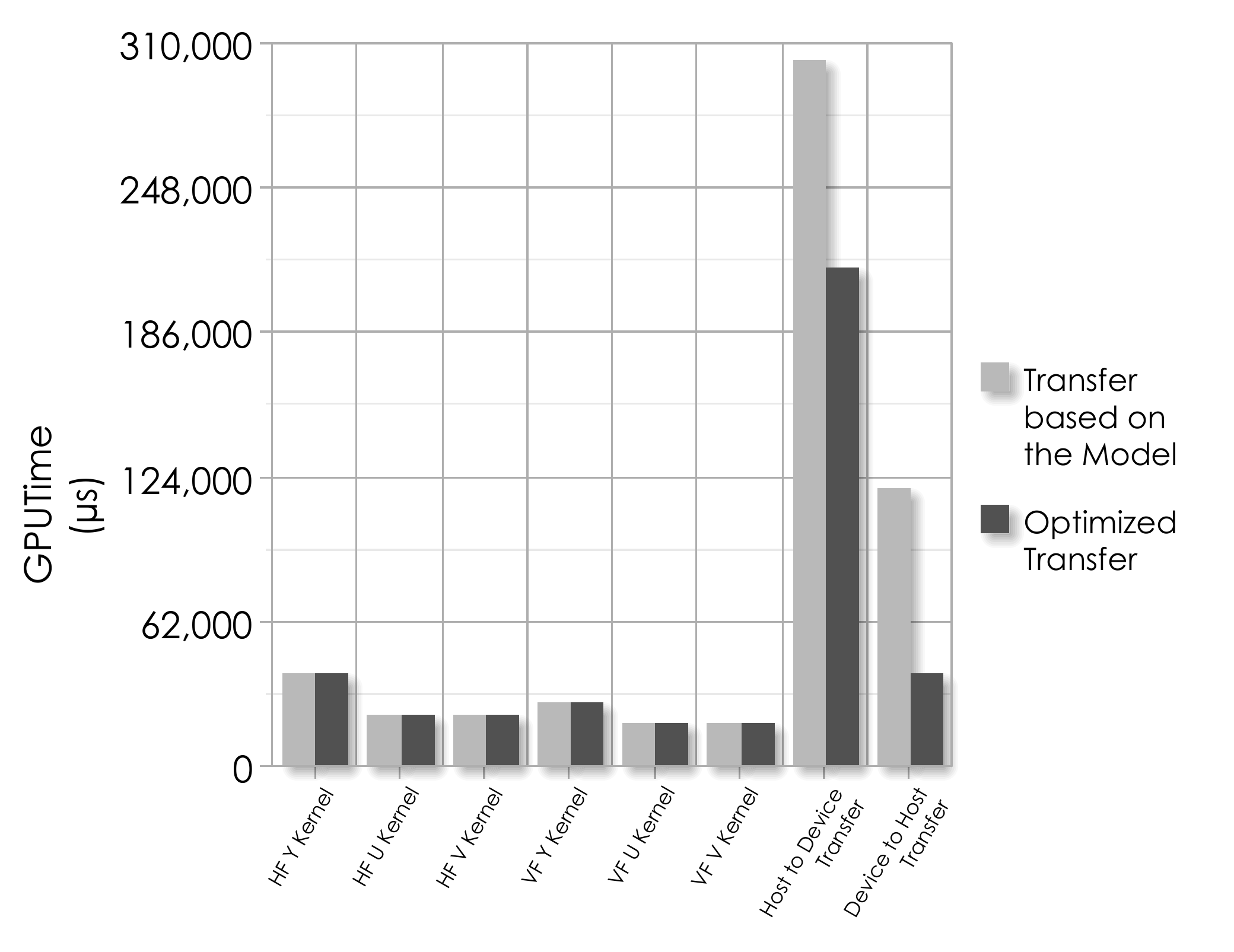}
\end{center}
\caption{Time Distribution on the GPU}
\label{fig-pizza}
\end{figure}

Four versions of the Downscaler were tested. The first one is a sequential version using the same structure defined in the figure \ref{fig-dscalerapp}. The other two versions are OpenCL automatically generated and the last one is a manually written OpenCL version. We have used a transformation chain that transforms model to model using QVTO \cite{qvt07} and model to text using Acceleo\cite{acceleo10}. The first OpenCL code is a not optimized program without any further analysis on memory transfers. The second one regards the memory transfers between host and device. Minimize these transfers reduces notably the total GPU execution time. As seen in the figure \ref{fig-pizza}, data transfers take a lot of time (more than 70\%) in the Downscaler application. The communication takes more time than computing process by the work-items. Time analysis in figure \ref{fig-pizza} demonstrates the bigger spent time in \emph{y-component} kernels due to their bigger topology and handled data. After the \emph{performance tuning}, no time changes occurs in kernels (as it was expected). However, we verify about 30\% and 70\% faster transfer times in \emph{host to device} and \emph{device to host} respectively.

\begin{figure}[!h]
\begin{center}
\includegraphics[width=.40\textwidth,height=.25\textwidth]{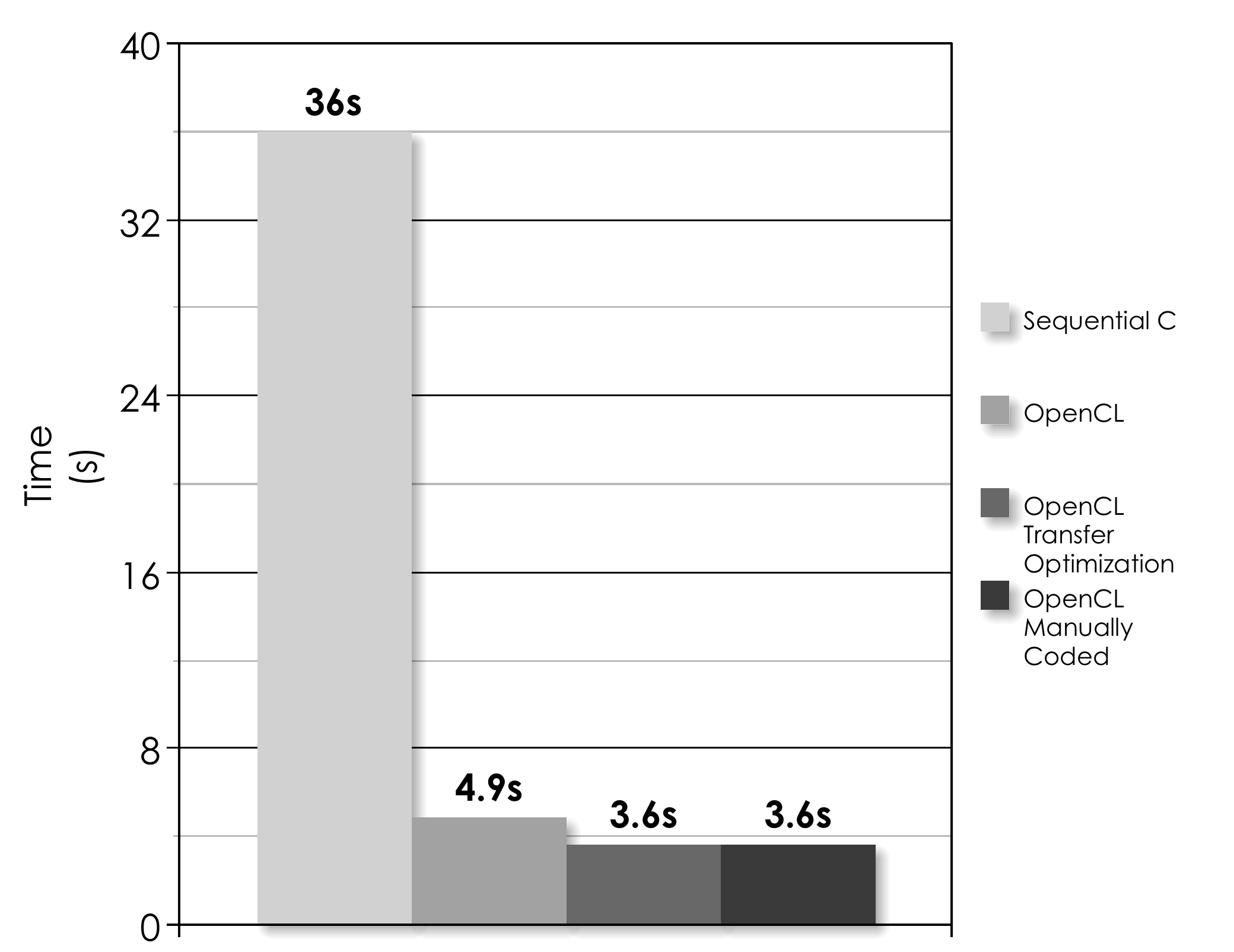}
\end{center}
\caption{Running Time Chart for the four Downscaler Implementations}
\label{fig-compresult}
\end{figure}

The figure \ref{fig-compresult} presents the total execution time of each implemented version. Both OpenCL codes give us good results with relation to CPU code. Using optimized transfers we can achieve speedups of 10x. In fact, structurally, the optimized version is really closer to manually written (considering the model designer is the code programmer). The decision of the topology and data transfers by the transformations (model compiler) were closely inspired by decisions taken if they would a human  programmer. For the optimized version we achieve about 25\% of speed-up. This is an enough expressive performance for two parallel implementations.

\section{Conclusion}
Even though we have used an application model not specially developed for GPU architectures, we have had good results at performance level. Parallel languages are hard to program and MDE approaches are well suitable to allow not specialized programmers creating parallel programs. The results presented in this work help us to certify the high potential of MARTE profile to create parallel applications for massively parallel processors. For the time being, we are implementing more optimization features in the transformation chain in order to ensure a stable and generic framework to create OpenCL applications exploiting, among other things, the memory hierarchy throughput.

\bibliographystyle{plain}
\bibliography{wsmbed.bib}

\end{document}